\documentclass[12pt]{article}

\pdfoutput=1
\usepackage{times}
\usepackage{graphicx}
\usepackage{hyperref}

\begin{document}

\title{Picasso, Matisse, or a Fake? \\ Automated Analysis of Drawings at the Stroke Level \\ for Attribution and Authentication\footnote{This paper is an extended version of a paper that will be published on the 32nd  AAAI conference on Artificial Intelligence, to be held in New Orleans, USA, February 2-7, 2018}}

\author{\href{mailto:elgammal@cs.rutgers.edu}{Ahmed Elgammal}$^{1,2} $ \footnote{
Corresponding author: {Ahmed Elgammal \href{mailto:elgammal@artrendex.com}{elgammal@artrendex.com}}} , Yan Kang$^{1}$, Milko Den Leeuw$^{3}$ \\
$^1$ \href{http://www.artrendex.com}{Artrendex LLC., NJ, USA}\footnote{\href{http://www.artrendex.com}{http://www.artrendex.com}} \\
$^2$ Department of Computer Science,
Rutgers University, NJ, USA\\
$^3$ Atelier for Restoration \& Research of Paintings (ARRS), \\The Hague, The Netherlands}


\maketitle

\begin{abstract}

This paper proposes a computational approach for analysis of strokes in line drawings by artists.  
We aim at developing an AI methodology that facilitates attribution of drawings of unknown authors in a way that is not easy to be deceived by forged art. 
The methodology used is based on quantifying the characteristics of individual strokes in drawings. We propose a novel algorithm for segmenting individual strokes. We designed and compared different hand-crafted and learned features for the task of quantifying stroke characteristics. We also propose and compare different classification methods at the drawing level. We experimented with a dataset of 300 digitized drawings with over 80 thousands strokes.  The collection mainly consisted of drawings of Pablo Picasso, Henry Matisse, and Egon Schiele, besides a small number of representative works of other artists. The experiments shows that the proposed methodology can classify individual strokes with accuracy 70\%-90\%, and aggregate over drawings with accuracy above 80\%, while being robust to be deceived  by fakes (with accuracy 100\% for detecting fakes in most settings).

\end{abstract}

\section{Introduction}

Attribution  of art works is  a very essential task for art experts.
Traditionally, stylistic analysis by expert human eye has been a main way to judge the authenticity of artworks. This has been pioneered and made a methodology by Giovanni Morelli (1816-1891) who was a physician and art collector, in what is known as Morellian analysis. This connoisseurship methodology relies on finding consistent detailed ``invariant'' stylistic characteristics in the artist's work that stay away from composition and subject matter.  For example Morelli paid great attention to how certain body parts, such as ears and hands are depicted in paintings by different artists, not surprisingly given his medical background. This methodology relies mainly on the human eye and expert knowledge. The work of van Dantzig~\cite{Pictology} that we follow in this paper belongs to this methodology. 

In contrast, technical analysis focuses on analyzing the surface of the painting, the underpainting, and/or the canvas material. There is a wide spectrum of imaging (e.g. infrared spectroscopy and x-ray), chemical analysis (e.g. Chromatography), and radiometric (e.g. carbon dating) techniques that have been developed for this purpose. Mostly, this analysis aims to get insights on the composition of the materials and pigments used in making the different layers of the work and how that relates to what materials, were available at the time of the original artist or what the artist typically used. These techniques are complementary and each of them has limitations to the scope of their applicability. We refer the reader to ~\cite{riederer1986detection} for comprehensive surveys of these techniques.

Analysis using computer vision and image processing techniques has been very sparsely and cautiously investigated in the domains of attribution and forgery detection (e.g.~\cite{guo2000off,johnson2008image,polatkan2009detection,li2012rhythmic}). Image processing has been used as a tool in conjunction with non-visual spectrum imaging, such as analysis of x-ray imaging to determine canvas material and thread count (e.g.~\cite{johnson2008image,liedtke2012canvas}).  

The question we address in this paper, is what role can the computer vision and AI technology plays in this domain given the spectrum of the other available technical analysis techniques, which might seem more conclusive.  We argue that developing this technology would complement other technical analysis techniques for three reasons.
First, computer vision can uniquely provide a quantifiable scientific way to approach the traditional stylistic analysis, even at the visual spectrum level. 
Second, it would provide alternative tools for the analysis of art works that lie out of the scope of applicability for the other techniques. For example, this can be very useful for detecting forgery of modern and contemporary art where the forger would have access to pigments and materials similar to what original artist had used). Third, computer vision has the potential to provide a cost-effective solution compared to the cost of other technical analysis methods. For example, in particular related to the topic of this paper, there are large volumes of drawings, prints, and sketches for sale and are relatively cheap (in the order of a few thousand dollars, or even few hundreds) compared to paintings. Performing sophisticated technical analysis in a laboratory would be more expensive than the price of the work itself.  This prohibitive cost makes it attractive for forgers to extensively target this market.


It is worthy to mention that several papers have addressed art style classification, where style is an art movement (e.g. Impressionism), or the style of a particular artist (e.g. the style of Van Gogh)~\cite{sablatnig,khan,Lombardi,Arora12,saleh2014toward}. Such stylistic analysis does not target authentication.  Such works use global features that mainly capture the composition of the painting. In fact, such algorithm will classify a painting done on the style of Van Gogh, for example, as Van Gogh, since it is designed to do so. 



\subsection*{Methodology}

\begin{figure}[tb]
\centering
 \includegraphics[width=4in]{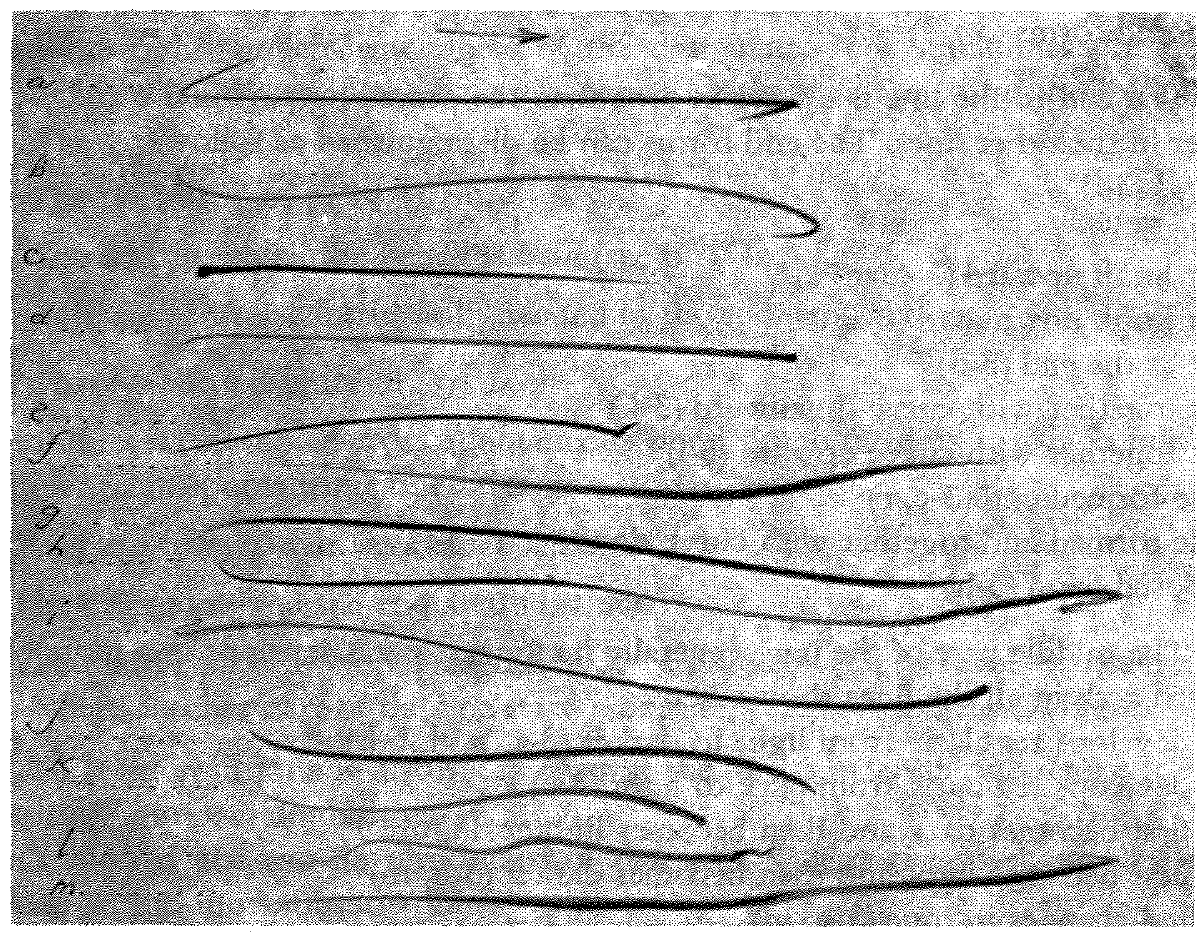}
\caption{Illustration of van Dantzig methodology on simple strokes. Spontaneous strokes differ in their shape and tone at their beginning, middle and end. Figure from~\cite{Pictology}}
\label{F:Dantzig}
\end{figure}

The methodology used in this paper is based on quantifying the characteristics of individual strokes in drawings and comparing these characteristics to a large number of  strokes by different artists using statistical inference and machine learning techniques.  This process is inspired by the Pictology methodology developed by  Maurits Michel van Dantzig~\cite{Pictology} (1903 - 1960).  Van Dantzig suggested several characteristics to distinguish the strokes of an artist, and suggested that such characteristics capture the spontaneity of how original art is being created, in contrast to the inhibitory nature of imitated art. 

Among the characteristics suggested by van Dantzig to distinguish the strokes of an artist are the shape, tone, and relative length of the beginning, middle and end of each stroke. The characteristics include also the length of the stroke relative to the depiction, direction, pressure, and several others. 
The list of characteristics suggested by van Danzig is comprehensive and includes, in some cases, over one hundred aspects that are designed for inspection by the human eye.
The main motivation is to characterize spontaneous strokes characterizing a certain artist from inhibited strokes, which are copied from original strokes to imitate the artist style. 

 In this paper we do not plan to implement the exact list of characteristics suggested by van Dantzig; instead we developed methods for quantification of strokes that are inspired by his methodology, trying to capture the same concepts in a way that is suitable to be quantified by the machine, is relevant to the digital domain, and facilitates statistical analysis of a large number of strokes by the machine rather than by human eye. 

We excluded using comparisons based on compositional and subject-matter-related patterns and elements.  Most forged art works are based on copying certain compositional and subject-matter-related elements and patterns. Using such elements might obviously and mistakenly connect a test subject work to figures and composition in an artist known works.
In contrast to subject matter and compositional elements, the characteristics  of individual strokes carry the artist's unintentional signature, which is hard to imitate or forge, even if the forger intends to do.

\subsection*{Contribution}

In this paper we propose   a computational approach for analysis of strokes in line drawings that is inspired and follow the principles of Pictology, as suggested by  van Dantzig.  We propose and validate a novel algorithm for segmenting individual strokes. We designed and compared different hand-crafted and learned deep neural network features for the task of quantifying stroke characteristics. We also propose and compare different classification methods at the drawing level. We experimented with a dataset of 300 digitized drawings with over 70 thousands strokes.  The collection mainly consisted of drawings of Pablo Picasso, Henry Matisse, and Egon Schiele, besides a small number of representative works of other artists. We extensively experimented on different settings of attributions to validate the proposed methodology. We also experimented with forged art works to validate the robustness of the proposed methodology and its potentials in authentication.

\begin{table*}[tbh]
\caption{Dataset collection: technique distribution}
\begin{center}
\scalebox{0.7}{
\begin{tabular}{|l|c|c|c|c|c|c|c|c|c|}
\hline
Technique & Pen/brush (ink) & Etching & Pencil & Drypoint & Lithograph & Crayon & Charcoal & Unknown & Total\\
\hline \hline  
Picasso &    80  &  38 &    8  &   2  &   2  &   0  & 0 & 0  & 130\\
Matisse &    45   & 10  &   5   &  2   & 14  &   1  & 0 &0 & 77\\
Schiele  &      0   &  0   & 10   &  0   &  0   &  5  & 4 & 17 & 36\\
Modigliani  & 0    & 0    & 9     & 0    &  0   & 8   & 1 &  0 & 18 \\
Others & 20  &   0  &   0  &   0  &   9  &   4  &   1 & 2  & 36 \\
\hline
Total    &    145  & 48   & 32   &  4  &  25   & 18  & 6 & 19  & 297\\
Strokes &     36,533     &  19,645     &   9,300     &    914    &    6,180      &  4,648      &   666  & 2,204 & 80,090 \\
\hline
\multicolumn{10}{|l|}{Others: Georges Braque, Antoine Bourdelle, Massimo Campigli, Marc Chagall, Marcel Gimond,} \\
\multicolumn{10}{|l|}{ Alexej Jawlensky, Henri Laurens, Andre Marchand, Albert Marquet, Andrà Masson, Andre Dunoyer Dr Segonzac, Louis Toughague} \\
\hline
\end{tabular}}
\end{center}
\label{T:technique}
\end{table*}%

\section{Detailed Methodology}
\subsection{Challenges}

The variability in drawing technique, paper type, size, digitization technology, spatial resolution, impose various challenges in developing techniques to quantify the characteristic of strokes that are invariant to these variability. Here we highlight some these challenges and how we addressed them.

Drawings are made using different techniques, materials and tools, including, but not limited to drawings using pencil, pen and ink, brush and ink, crayon, charcoal, chalk, and graphite drawings. Different printing techniques also are used such as etching, lithograph, linocuts, wood cuts, and others. Each of these techniques results in different stroke characteristics. This suggests developing technique-specific models of strokes. However, typically each artist prefers certain techniques over others, which introduce unbalance in the data collection, which need to be addressed. Therefore, in this paper we are testing two hypotheses: technique specific vs. across technique comparisons, to test if we can capture invariant stroke characteristic for each artist that persists across techniques.  

Drawings are executed on different types of papers, which, along with differences in digitization, imply variations in the tone and color of the background. This introduces a bias in the data. We want to make sure that we identify artists based on their strokes and not based on the color tone of the paper used. Different types of papers along with the type of ink used result in different diffusion of ink at the boundaries of the strokes which, combined with digitization effects, alter the shape of the boundary of the stroke.

Drawings are made on different-sized papers, and digitized using different resolutions. The size of the original drawing as well as the digitization resolution are necessary to quantify characteristics related to the width or length of strokes.  Therefore, in this paper we quantify the characteristics of the strokes in a metric basis after converting all the measurements to the metric system.


\subsection{Data collection}

A collection of 297 drawings were gathered from different sources to train, optimize, validate, and test the various classification methodologies used in this study. The drawings selected are restricted to line drawings, i.e, it excludes drawings that have heavy shading, hatching and water-colored strokes. The collection included drawings and prints by Picasso (130), Henry Matisse (77), Egon Schiele (36), Amedeo Modigliani (18), and a small representative works of other artists (36), ranging from 1910-1950AD. These artists were chosen since they were prolific in producing line drawings during the first half of the Twentieth century.  
 
The collection included a variety of techniques including: pen and ink, pencil, crayon, and graphite drawings as well as etching and lithograph prints. Table~\ref{T:technique} shows the number of drawings for each artist and technique. In the domain of drawing analysis it is very hard to obtain a dataset that is uniformly sampling artists and techniques. The collection is biased towards ink drawings, executed mostly with pen, or using brush in a few cases. There is a total of 145 ink drawings in the collection.  The collection contains more works by Picasso than other artists. In all the validation and test experiments an equal number of strokes were sampled from each artist to eliminate data bias.

\begin{figure}[htb]
\centering
 \includegraphics[width=5in]{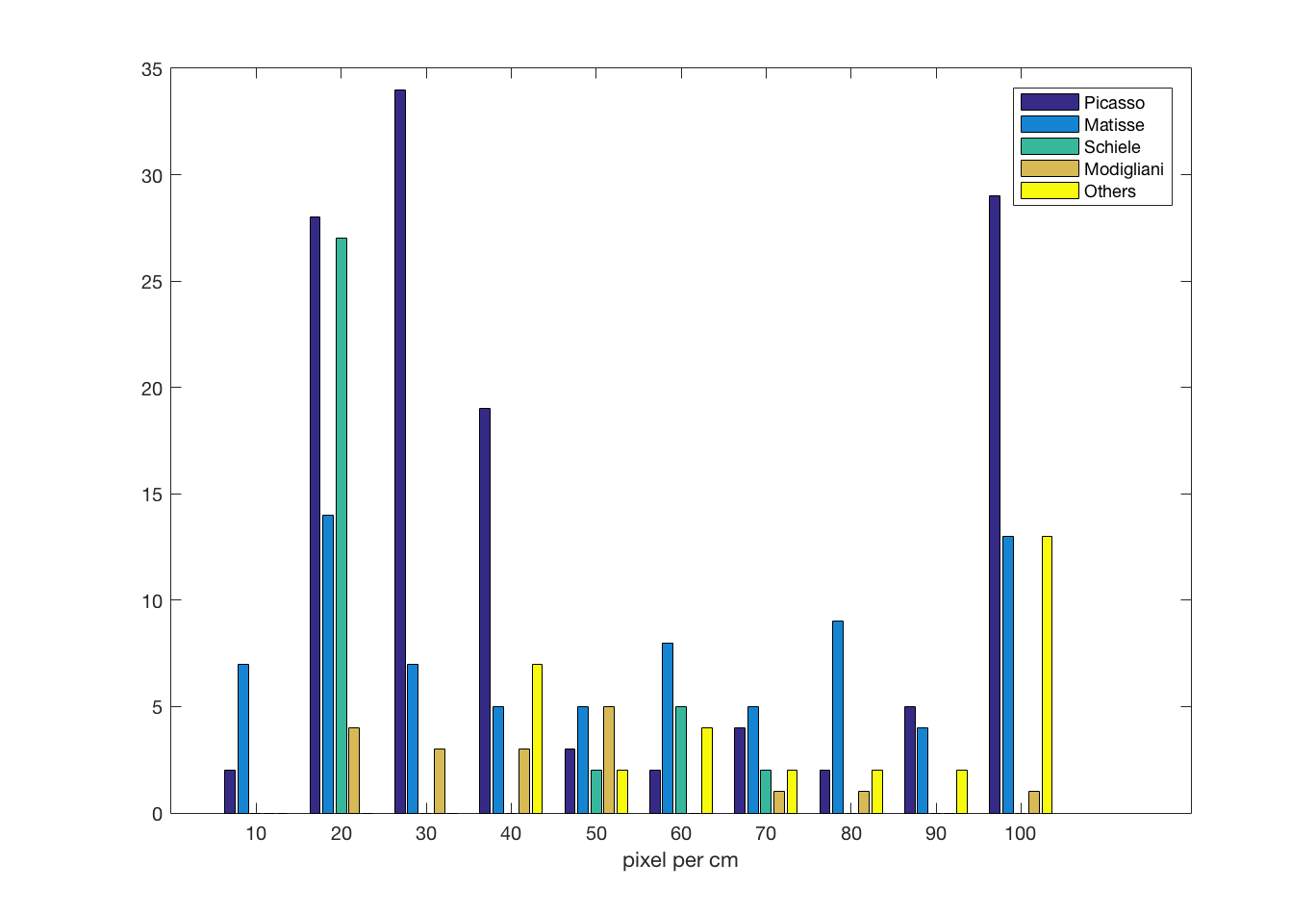}
\caption{Distribution of digitization resolution (in Pixel per cm units)}
\label{F:resolution}
\end{figure}



The collection included digitized works from books, downloaded digitized images from different sources, and screen captured images for cases where downloading was not permitted. The resolution of the collected images varies depending on the sources. The effective resolution varies from 10 to 173 pixel per cm depending on the actual drawing size and the digitized image resolution. Figure~\ref{F:resolution} shows the distribution of the digitized images resolution.  Given this wide range of resolutions, the algorithms and features used were designed to be invariant to the digitization resolution.

\noindent{\em Fake drawing dataset:} In order to validate the robustness of the proposed approaches against being deceived by forged art, we commissioned five artists to make drawings similar to those of Picasso (24), Matisse (39) and Schiele (20) using the same techniques. We collected a total of 83 drawings (24, 39, 20). None of these fake drawings was used in training the models. We only used them for testing. 

\begin{figure}[pth]
\centering
\includegraphics[width=6.5in]{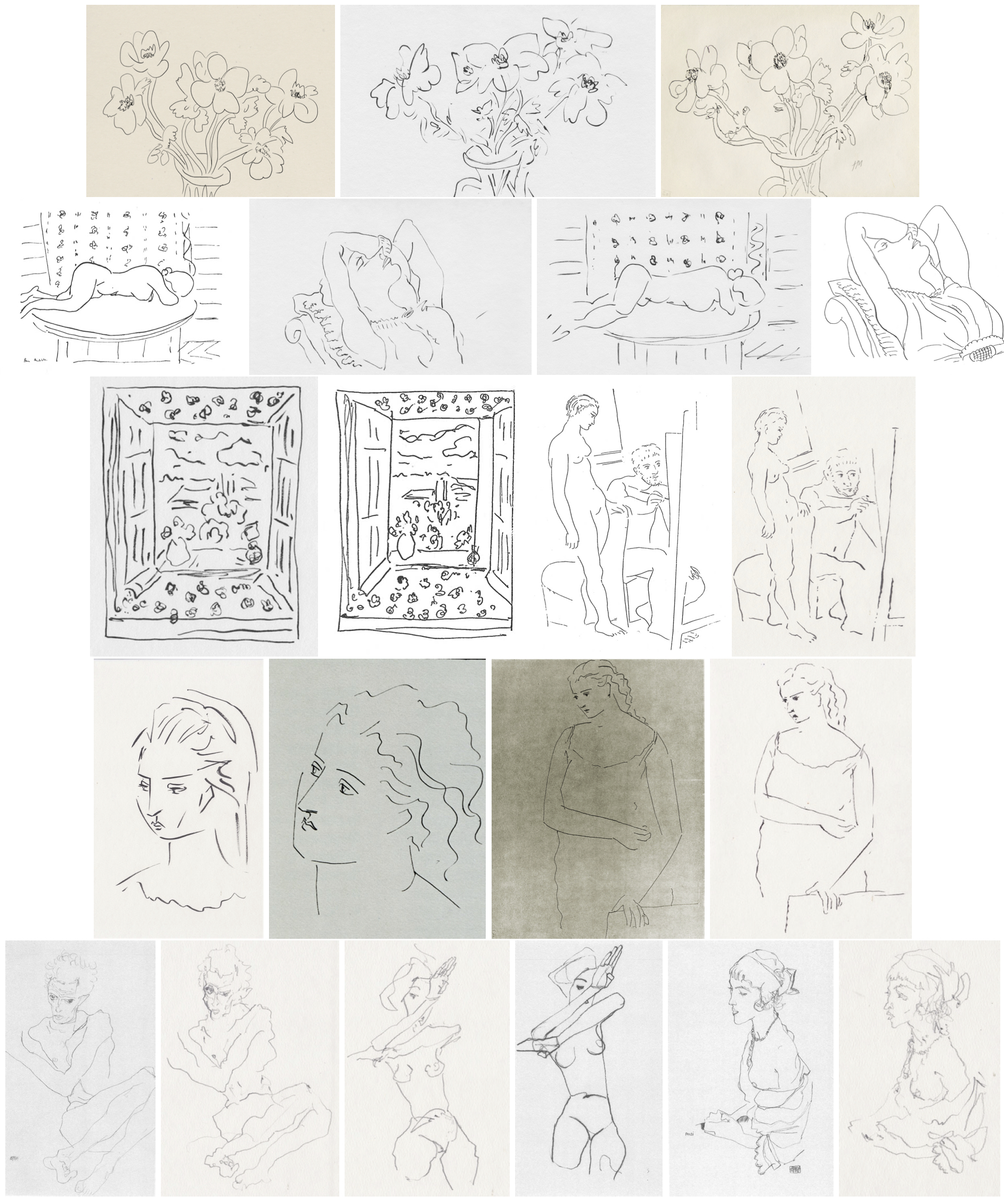}

\caption{Examples of images of the fake dataset mixed up with real images of drawings by Matisse, Picasso, and Schiele. See the key at the end of the document to tell which are real and which are fake!}
\label{F:Fakes}
\end{figure}

Figure~\ref{F:Fakes} shows examples of the fake dataset mixed up with real drawings. 

Because we do not expect the reader to be experts in authentication in art, to be able to judge the quality of the fake drawings in isolation, we deliberately mixed up a collection of the fake drawings with real drawings in Figure~\ref{F:Fakes}. If the reader is interested to know which of these images are of fake or real drawings, please refer to the end of the paper.


\section{Stroke Segmentation}

\begin{figure}[ph]
\centering
\includegraphics[width=5.7in]{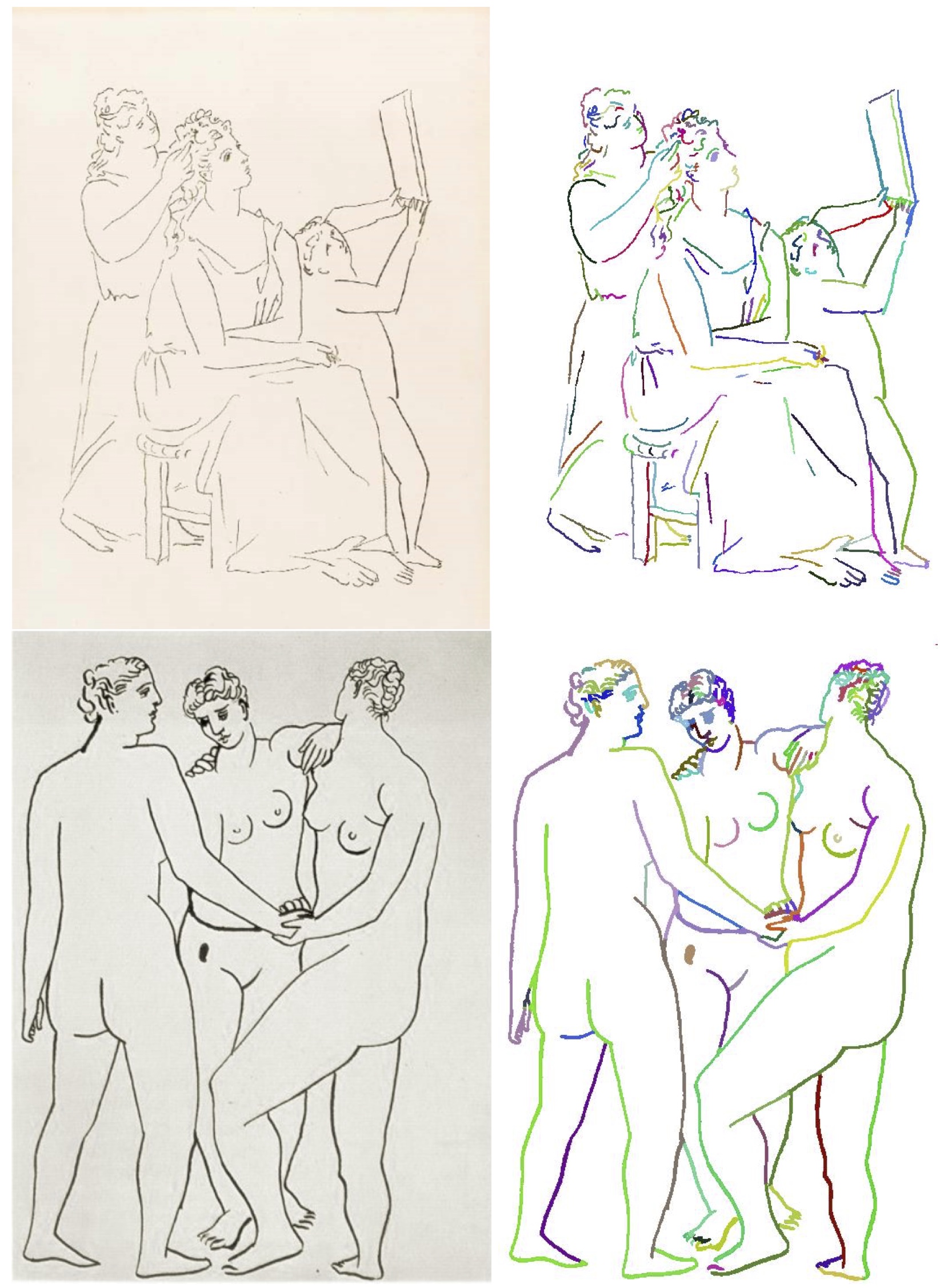}
 
\caption{Examples of segmentation results. Top: Picasso lithograph. Bottom: Picasso ink drawing. - best seen in color }
\label{F:Segmentation}
\end{figure}

\begin{figure*}[thp]
\centering
\includegraphics[width=5.7in]{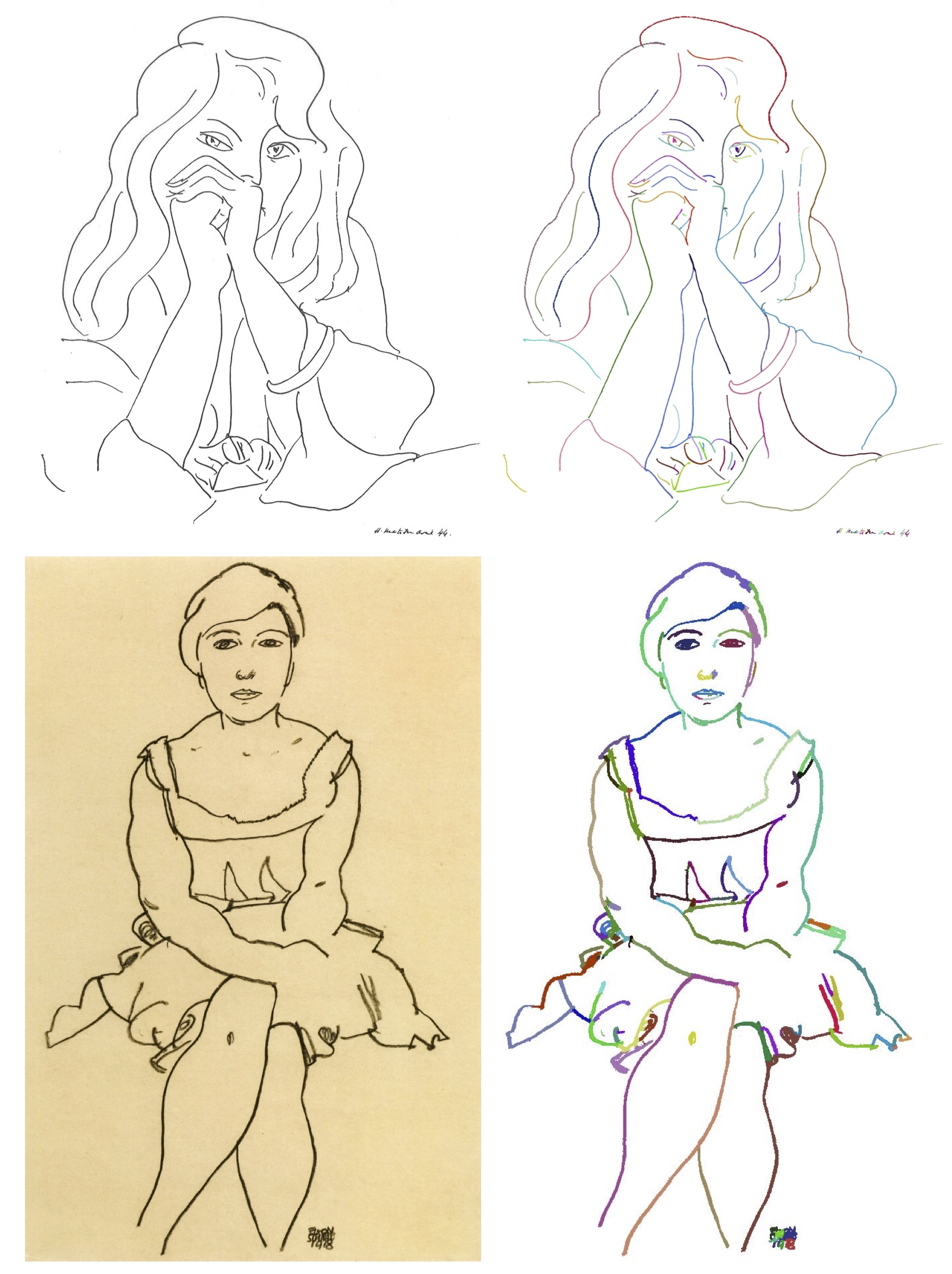}
\caption{Examples of segmentation results. Top: Matisse etching. Bottom: Schiele ink drawing. - best seen in color }
\label{F:Segmentation2}
\end{figure*}

The stroke segmentation algorithm will be described here. See sample results in Figure~\ref{F:Segmentation} and~\ref{F:Segmentation2}.

A typical isolated stroke is a line or curve, with a starting point and endpoint. A stroke can have zero endpoints (closed curve) or 1 endpoint, which are special cases that do not need further segmentation. However, strokes typically intersect to form a network of tangled strokes that needs to be untangled. A network of strokes is characterized by having more than 2 endpoints. Since strokes are thin elongated structures; a skeleton representation would preserve their topological structure even in a network configuration~\cite{lam1992thinning}. Therefore, the segmentation of strokes is done on such a skeleton representation.

There is a large classical literature in computer vision on detecting junctions on edge maps as a way to characterize object boundaries, infer about three-dimensional structure and form representations for recognition. 
Unlike classical literature which look at natural images, in our case detecting junctions and endpoints is fortunately relatively easy since they persists in a skeleton representation of the network of strokes.  On the other hand, the challenge in our case is to use the information of such junctions and endpoints to segment individual strokes. 

In our case, junctions play crucial role in identifying the intersections between strokes. There are two basic ways strokes intersect: occluder-occluded configuration to form a T-junction or two strokes crossing each other to form an X-junction.  A T-junction is a continuation point of the occluding stroke and an endpoint for the occluded stroke. We need to preserve the continuation of the occluding stroke at the T-junction.

The stroke segmentation algorithm takes a network of strokes and identifies one occluding stroke at a time and remove it from the network of strokes to form a residual network(s) that is recursively segmented. This is achieved by constructing a fully connected graph whose vertices are the endpoints in the network and edges are weighted by the cost of reaching between each two endpoints. The cost between two endpoints reflects the bending energy required at the junctions. 

Let the endpoints in a network of strokes denoted by $e_1,\cdots,e_m$ and let the junction locations denoted by $j_1,\cdots,j_n$. 
The cost of the path between any two end points $e_i$ to $e_j$ is cumulative curvature along the skeleton path between them, where the curvature is only counted close to junctions. The rational is that it does not matter how much bending a stroke would take as long as it is not at junctions.
Let $\gamma(t): [0:1] \rightarrow R^2$ be the parametric representation of the skeleton curve connecting $e_i$ and $e_j$. The cost is defined as
\[ c(e_i,e_j) = \int_{0}^{1} \kappa(t) \cdot \phi(\gamma(t)) dt\]
where $\kappa(\cdot) $ is the curvature and  $\phi(\cdot)$ is a junction potential function, which is a function of the proximity to junction locations defined as
\[ \phi(x) = \frac{1}{n} \sum_{i=1}^n e^{(x-j_i)^2/\sigma} \]

After the graph construction, the minimum cost edge represents a path between two endpoints with minimum bending at the junctions, which corresponding to an occluding stroke. In case of a tie, the path with the longest length is chosen. The optimal stroke is removed from the skeleton representation and from the graph. This involves reconnecting the skeleton at X-junctions (to allow the detection of the crossing strokes) and new endpoints have to be added at T-junctions (to allow the detection of occluded strokes. Removing a stroke from the graph involves removing all edges corresponding to paths that go through the removed stroke. This results in breaking the graph to one or more residual subgraphs, which are processed recursively.

%

\section{Stroke Analysis Methodology}
\label{S:StrokeAnalysis}
\subsection{Quantifying Stroke Characteristics}

This section explains the process of quantifying the characteristics of individual strokes and the extracted features used to represent each stroke. 
The goal is to construct a joint feature space that captures the correlation between the shape of the stroke, its thickness variation, tone variation, local curvature variation.  For this purpose we studied two different types of features and their combination: 1) Hand-crafted features capturing the shape of each stroke and its boundary statistics, 2) Learned-representation features capturing the tone variation as well as local shape characteristics.  The next two subsection describe each of these features.


\subsubsection{Hand-crafted Features}

In our study, each stroke is represented by its skeleton, its boundary, and the rib length around the skeleton. The following descriptors are extracted to quantify the characteristics of each stroke. All the descriptors are designed to be invariant to translation, rotation, scaling, and change in digitization resolution.

\noindent{\em Shape of the boundary:} 
The shape of the stroke boundary is quantified by Fourier descriptors~\cite{Burger2016FD}. Fourier descriptors are widely used shape features for a variety of computer vision applications such as character recognition and shape matching. Fourier descriptors provide shape features that are proven to be invariant to translation, scaling, rotation, sampling, and contour starting points~\cite{Burger2016FD}. 
We used 40 amplitude coefficients (first 20 harmonics in each direction) to represent the shape of the boundary of the stroke.   

\noindent{\em Reconstruction error profile:} 
The mean reconstruction error, as a function of the number of harmonics used to approximate the shape of the strokes, is used as a descriptor of the smoothness of the contour and the negative space associated with the stroke. 
In particular, we compute the mean reconstruction error at each step while incrementally adding more harmonics to approximate the shape of the stroke.  The reconstruction error profile is normalized by dividing by the stroke mean width in pixels to obtain a descriptor invariant to digitization resolution. 

 \noindent{\em Contour Curvature descriptor:} 
 To quantify the curvature of the stroke contours, we use the first and second derivatives of the angular contour representation. The distributions of these derivatives are represented by their histograms. 
 
 \noindent{\em Stroke thickness profile:} 
 To quantify the thickness of the stroke, we compute the mean and standard deviation of the rib length around the skeleton of the stroke, as well as a histogram of the rib length.  All rib length measurements  are mapped to mm units to avoid variations in digitization resolution. 

\noindent{\em Stroke Length:} 
The length of the stroke is quantified as the ratio between the stroke skeleton length to the canvas diagonal length. This measure is invariant to digitization resolution. 


\subsubsection{Deep Learned Features using RNNs}

\noindent{\em GRU Classification with Truncated Back Propagation Through Time:}

Other than the traditional feed-forward neural networks specialized at fixed size input, e.g. images, recurrent neural network (RNN) could handle variable length sequence input $x = (\textbf{x}_1, \cdots, \textbf{x}_T)$ and either fixed length output or variable length output $y = (\textbf{y}_1, \cdots, \textbf{y}_T)$ by utilizing the hidden state within. RNN sequentially takes input $\textbf{x}_t$ from the input sequence and update its hidden state by:
$$h_t=\phi_\theta(h_{t-1}, \textbf{x}_t)$$
$\phi_\theta$ is a nonlinear activation function with $\theta$ as parameters. In each time step, a corresponding output could be generated through:
$$\hat{y}_t = g_\theta(h_t, x_t)$$
$g_\theta$ is an arbitrary parametric function that is trained together with the recurrent cell.\\
Recently, it has been widely shown that the more complicated RNN model such as Long Short-Term Memory (LSTM, \cite{LSTM}) or Gated Recurrent Unit (GRU, \cite{GRU2}) would eliminate the problem of vanishing gradient \cite{vanishinggradient1}, \cite{vanishinggradient2}. LSTM and GRU introduce some gating units that can automatically determine how much the information flow could be used in each time step, by which the vanishing gradient could be avoided.

In GRU, two gate units are introduced: reset gate and update gate. Reset gate controls how much former hidden state would be used in calculating the current time step candidate $\hat{h_t}$.
$$r_t = \sigma(U_rh_{t-1} + W_rx_t)$$
Update gate controls how much current candidate $\hat{h_t}$ would be updated and how much old hidden state would be kept.
$$z_t = \sigma(U_zh_{t-1} + W_zx_t )$$
Then the candidate hidden state $\hat{h_t}$ would be:
$$\hat{h_t} = tanh(U(r_t\odot h_{t-1}) + Wx_t)$$ 
And the final hidden state $h_t$ at time $t$ is updated though:
$$h_t = z_t\odot \hat{h_t} + (1-z_t)\odot h_{t-1}$$ 
Where U, W, Ur, Wr, Uu, Wu are the learned weight matrices and the biases are omitted here to make the equations more compact. $\odot$ is a pointwise multiplication, and $\sigma$ is a sigmoid function.

Given a stroke, a sequence of patches of fixed size are collected along the skeleton of the stroke and fed to a GRU model as inputs. We tested both fixed size patches or adaptive size patches where the radius of the patch is a function of the average stroke width in the drawing. In both cases the input patches are scaled to 11x11 input matrices. To achieve invariant to the direction of the stroke, each stroke is sampled in both directions as two separate data sequences (at classification, both a stroke and its reverse either appear in training or testing splits ).
We normalized the grey scale into range (-1, 1), and flattened the 11 $\times$ 11 image into a 121-dimension vector. The activation function we used in experiments is $tanh$ function. Parameters are initialized from normal distribution with mean = 0, standard deviation = 1. After comparing several optimizer functions, we found that the RMSProp optimizer with learning rate 0.001 outperforms others.

The gradient is globally clipped to be less than 5 to prevent from gradient exploding,. And to avoid  gradient vanishing, we calculated the gradient by the truncated Back Propagation Through Time. Each sequence is unrolled into a fixed size $\tau$ steps ($\tau = 30$ in the experiments) at each time to calculate the gradient and to update the network's parameters. The label of original sequence is assigned to each unrolling. Between each unrolling, the hidden state is passed on to carry former time steps information. And within each unrolling, only the last time step hidden state is used in the final linear transformation and Softmax function to get the predicted score of each class. The loss function we used was the cross entropy.\\
$$x_\tau = (\textbf{x}_t, ... \textbf{x}_{t+\tau})$$ 
$$h_{t+\tau} = GRU(h_{initial}, x_\tau)$$
$$\hat{y} = softmax(U_sh_{t+\tau})$$
$$loss = -\sum{ylog(\hat{y})}$$
$$h_{initial} = h_{t+\tau}$$

%
%
%

\subsection{Stroke  Classification}

For the case of hand-crafted features, strokes are classified using a support vector machine (SVM) classifier~\cite{Cortes1995}.  We evaluated SVM using Radial basis kernels as well as polynomial kernels. The classifier produces posterior distribution over the classes. For the case of learned GRU features, the classification of strokes is directly given by the trained networks. SVM was used to combine hand-crafted features with the learned features in one classification framework. In such case, the activation of the hidden units were used as features, and combined to the hand-crafted features.

\subsection{ Drawing classification:} 

A given drawing is classified by aggregating the outcomes of the classification of its strokes. We used four different strategies for aggregating the stroke classification results, as described below.

\begin{itemize}
\item {\em Majority Voting:}  In this strategy each stroke votes for one class. All strokes have equal votes regardless of the certainty of the output of the stroke classifier.

\item {\em Posterior aggregate:} In this strategy each stroke votes with a weight equal to its posterior class probability. 
This results in reducing the effect of strokes that are not classified with high certainty by the stroke classifier.

\item {\em k-certain voting:} In this strategy, only the strokes with class posterior greater than a threshold $k$  are allowed to vote. This eliminates effect of uncertain strokes. 

\item {\em certainty weighted voting: } In this strategy each stroke vote is weighted using a gamma function based on the certainty of the stroke classifier in classifying it.  

\end{itemize}

\section{Validation Experiments}
\noindent This section describes the experiments conducted to test and validate the performance of the proposed stroke segmentation, stroke classification, and the drawing classification approaches on the collected dataset. In particular, the experiments are designed to test the ability of the algorithms to determine the attribution of a given art work and test its robustness to forged art.

\subsection{Segmentation Validation}
\begin{figure}[tbph]
\centering
  \includegraphics[width=6.5in]{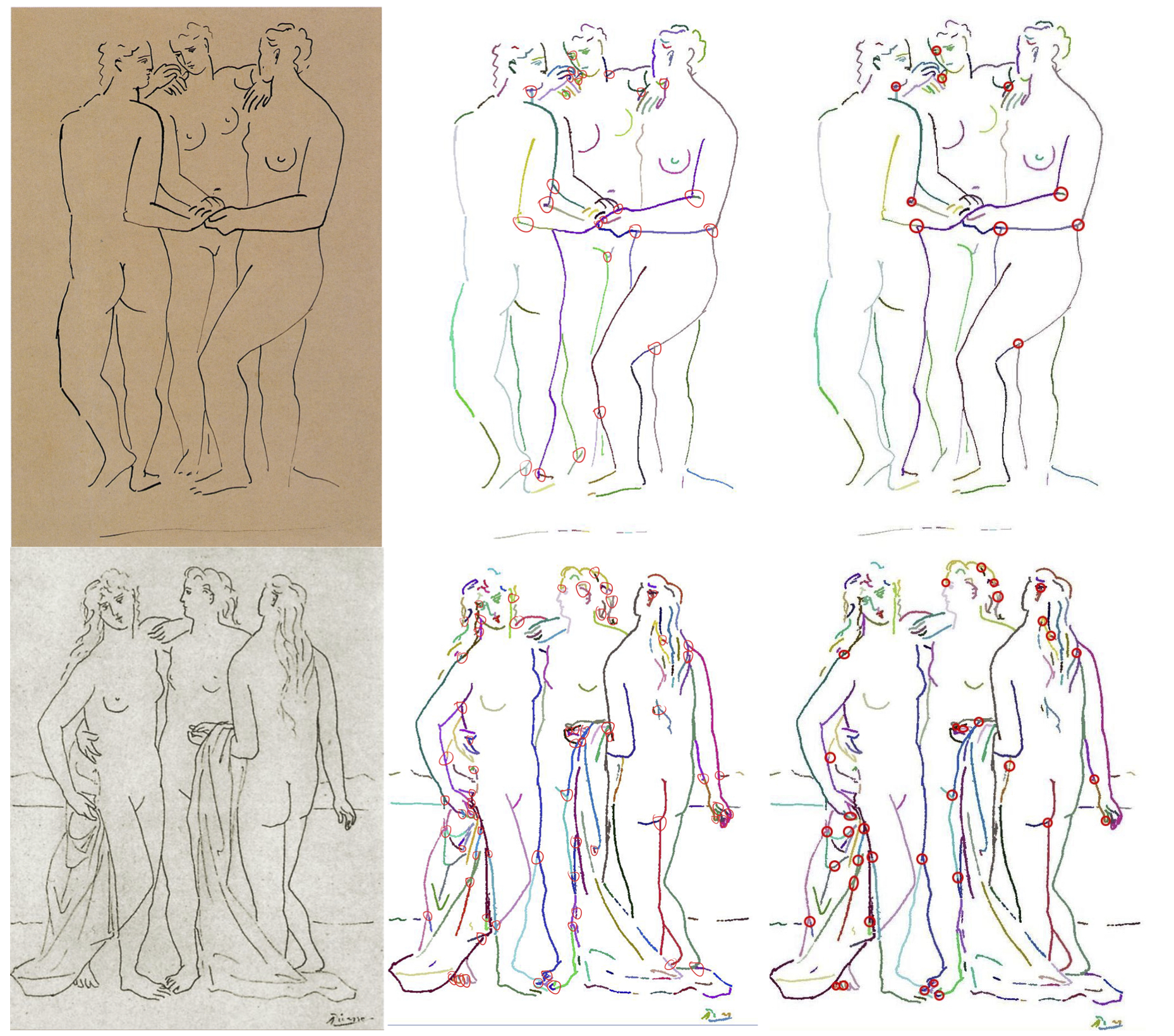}
\caption{Example of two drawings by Picasso, stroke segmentation results, and segmentation errors marked by two artists.}
\label{F:SegValidate}
\end{figure}

Validating the segmentation algorithm is quite challenging since there is no available ground truth segmentation and because of the difficulty of collecting such annotation. It is quit a tedious process for a human to trace individual strokes to provide segmentation of them, specially such task requires certain level of expertise. To validate the segmentation algorithm, we collected 14 drawings with medium difficulty (in terms of number of strokes) from the collection and showed the segmentation results to two artists and asked them independently to locate errors in the  segmentations. Figure~\ref{F:SegValidate} shows an example of a drawing with its two annotations of the results. A closer look highlights that annotators make several mistakes (false positive, and false negatives).  
Table~\ref{T:SegmentationValid} shows the number of marked errors for each sample image by two evaluators.  The overall error per annotator is computed as: 
Error rate = total marked errors at junctions / total number of strokes;
where the total is aggregated over all evaluated images. The average error rate over the two annotators is 12.94\% , counting all labeled errors by annotators. 
The annotation shows large deviations between the two annotators, with mean deviation 24.93 and standard deviation 12\%. This highlight the challenge in validating the segmentation results by human annotation. However, most of the marked errors are at small detailed strokes that are hard to segment, even by the human eye, and does not contribute much to the classification of strokes since small strokes are filtered out anyway. 

\begin{table}[htbp]
\caption{Validation of Stroke Segmentation}
\begin{center}
\scalebox{0.7}{
\begin{tabular}{|l|l|l|l|l|}
\hline
Sample & Number of & 	Evaluator 1 &	Evaluator 2	& Absolute deviation  \\
             &  Strokes     &   Marked Errors & Marked Errors & between Evaluators \\
\hline
1 & 596 &	34	&75 &	41 \\
2 & 366 &	37	&17	&20 \\
3 &314 &	38	&11	&27 \\
4 & 216 &	24	&11	&13 \\
5 & 267 &	69	&13	&56 \\
6 & 122 &	40	&14	&26 \\
7 &136 &	28	&10	&18 \\
8 &131 &	32	&12	&20 \\
9 &102 &	22	&10	&12 \\
10& 71	& 17	&6	&11 \\
11 &159 &	48	&15	&33 \\
12 &123 &	30	&8	&22 \\
13 & 103 &	25	&10	&15 \\
14 &196 &	65	&30	&35 \\
\hline 
Total & 2902 &	509	& 242	& \\
Mean &    &   &    & 24.93 \\
Std & & &	 &		12.72 \\
\hline
\end{tabular}
}
\end{center}
\label{T:SegmentationValid}
\end{table}

\subsection{Stroke Classification}
{\em Evaluation Methodology:} In all experiments the image datasets were split into five 80/20\% folds to perform five-fold cross validation. Since strokes from the same drawings might share similar characteristics, we did this splits at the image level and not at the stroke level. For each fold, after splitting the images to train and test sets, equal number of strokes were sampled for each artist class for training and testing to avoid the bias in the data, which is significant in our case. We evaluated different classification settings including pair-wise classification, and one-vs-all classification, and multi-class classification. Extensive ablation  studies are also performed to evaluate the different features and their effects, as well as to choose the optimal settings.

\subsubsection{Stroke Classification Validation - Technique Specific - Pairwise:} 
For testing technique-specific classifiers, we trained pairwise classifiers to discriminate between Picasso and Matisse drawings made using either pen/ink or etching. We chose these two techniques and these two  artists since they have the largest representation in our collection. Table~\ref{T:StrokePairWise} shows the stroke classification results. The experiments is done using five-fold cross validation and the mean and standard deviations are reported. The table shows a comparison between the different types of proposed features.

\begin{table}[htp]
\caption{Validation of Stroke Classifier - Technique specific (Picasso vs Matisse)}
\begin{center}
\scalebox{0.9}{
\begin{tabular}{|l|l|l|}
\hline
 \multicolumn{3}{|c|}{Ink Drawing (Pen/Brush)  (Picasso vs Matisse)}\\
 \hline
Approach & Train & Test \\
\hline
Hand-Crafted - SVM-RBF & 87.99\% (0.39\%)	&  79.16\% (0.26\%) \\
Hand-Crafted - SVM-POLY & 79.88\% (0.14\%)& 77.17\% (0.58\%)\\
GRU  					 &  84.92 \% (  1.89\%) &  65.86  (13.58 \% )  \\
\hline
 \multicolumn{3}{|c|}{Etching Prints  (Picasso vs Matisse)}\\ 
\hline
Approach & Train & Test \\
\hline
Hand-Crafted - SVM-RBF & 94.53\% (0.22\%) &  84.18\% ( 0.85 \%) \\
Hand-Crafted - SVM-POLY &  94.27\% (0.21\%) &  93.09\% ( 0.88\%) \\
GRU						&  83.74\% (4.60 \% ) &	75.08\% (8.11\%) \\
\hline
\end{tabular}
}
\end{center}
\label{T:StrokePairWise}
\end{table}

\subsubsection{Stroke Classification Validation - One-vs-all:}

In this experiment a one-vs-all classification settings is used to build classifiers for Picasso-vs-Non-Picasso, Matisse-vs-Non-Matisse, Schiele-vs-Non-Schiele. These three artists are chosen since they have enough data for training and testing the classifiers in a five-fold split setting. The classifiers are then evaluated on the fake dataset (see Section~\ref{S:ValidateDraw} ). 

We evaluated the performance of two settings: 

\begin{enumerate}
\item across-techniques: we evaluated the performance of the stroke classifiers on all techniques combined to evaluate whether the classifier can capture an invariant for the artist regardless of the technique used. 

\item Technique-specific: in this setting each classifier is trained and tested using strokes from the same drawing technique. Given the data collection, we tested a) Picasso-vs-Non-Picasso classifier using ink/pen, b) Matisse-vs-Non-Matisse classifier using ink/pen, c) Schiele-vs-Non-Schiele using pencil. 
\end{enumerate}

Table~\ref{T:Stroke1vsAll} shows the mean and standard deviations of the five folds for the hand-crafter features, the GRU features and the combination. Both types of features have very good stroke classification performance. GRU has better performance over the three artists tested. Combining the features further improved the results and reduced the cross-fold variances, which indicate that both types of features are complementary to each other as we hypothesized.

Comparing the performance of stroke classifiers on both the technique-specific and across-technique settings, we notice that in both cases the classifiers performed well. The GRU performed better in the across-technique settings than in the technique-specific setting, which can be justified by the lack of data in the later case. 

\begin{table*}[tbh]
\caption{Validation of Stroke Classifiers - One-vs-All }
\begin{center}
\scalebox{0.64}{
\begin{tabular}{|l|l|l|l|l|l|l|l|}
\hline
\multicolumn{8}{|c|}{ {\bf Across-Techniques} - Mean (std) of five folds} \\
\hline
 & & \multicolumn{2}{|c|}{Hand-crafted + SVM} & \multicolumn{2}{|c|}{GRU} & \multicolumn{2}{|c|}{Combined} \\
 \hline
Classifier & technique &train & test & train & test & train & test \\
\hline
Picasso vs. all & All & 72.59\% (1.19 \%) & 67.26\% (8.37 \%) & 81.92\% (2.59 \%) & 75.09\% ( 5.09\%) & 86.05\% (1.08 \%) & 78.54\% (4.36 \%) \\
 Matisse vs. all & All & 65.83\% (1.72 \%) & 60.61\% ( 8.71\%) & 81.01\% (3.41 \%) & 72.68\% (5.58 \%) & 87.92\% ( 1.73\%) & 77.08\% ( 4.33\%) \\
Schiele vs. all & All & 84.76\% ( 0.91\%)  & 81.49\% (3.30 \%)  & 85.55\% (1.74 \%)  & 78.54\% ( 8.77\%)  & 91.85\% (0.87 \%)  &86.20 \% (3.78\%)  \\
\hline
\hline
\multicolumn{8}{|c|}{ {\bf Technique-specific} - Mean (std) of five folds} \\
\hline
 & &\multicolumn{2}{|c|}{Hand-crafted + SVM} & \multicolumn{2}{|c|}{GRU} & \multicolumn{2}{|c|}{Combined} \\
 \hline
Classifier & Technique & train & test & train & test & train & test \\
\hline
Picasso vs. all & Pen/Ink & 73.20\% 	( 2.21\%)  & 68.93\% ( 7.04\%)  & 	84.08\% ( 2.20\%) & 	72.24\% (1.87 \%)  &  88.40\% ( 1.19\%) & 75.92\% ( 4.22\%)  \\
Matisse vs. all & Pen/Ink & 73.35\%	 ( 1.99 \%) &  70.08\% ( 7.94\%) &	86.88\% (1.98 \%) &	 75.03\% ( 6.47\%) &	 91.56\% ( 1.03\%) &	 79.10\% (6.65 \%)  \\
Schiele vs. all & Pencil & 82.58\%	( 2.78\%)  & 75.39\%	 (20.64 \%) &   94.33\%  (3.52 \%) & 69.60\% (20.62 \%) & 91.30\% (4.57 \%) &	72.93\%	(19.67 \%)  \\
\hline
\end{tabular}
}
\end{center}
\label{T:Stroke1vsAll}
\end{table*}

\subsubsection{Multi-class Stroke Classification Experiment}
Although for attribution and authentication one-vs-all setting is the most obvious choice, we also tested a multi-class setting for completeness. In this experiment we train and test stroke classifiers built to discriminate between 5 classes: Picasso, Matisse, Schiele, Modigliani, and Others. The challenge in this setting is that training and test data will be bounded to the class that has the least number of samples (since we equalize the number of samples in training and test sets to avoid data bias). In this experiment we compared the performance of the hand-crafted features and the GRU features.  For the GRU, the output directly has 5 nodes to encode the classes. For the hand-crafted features we used error-correcting output codes (ECOC) classification setup where binary SVM classifiers were trained for each pair of classes. 

Table~\ref{T:MultiClass} shows the results of five-fold cross-validation experiment. There is a significant difference in performance between the two types of features in this experiment, which is far from the differences in all other experiments. We hypothesize that this is because the ECOC setting limits significantly the  number of data samples used for training each binary classifier, while the GRU utilized the samples from all 5 classes in training. In particular, the number of training strokes per classes in this experiment were 1418, 1656, 1551, 1162, 1317  for each of the 5-fold splits respectively, which is a very small number. As a conclusion, we rule out the use of multi-class setting for attribution and authentication due to the hardship in obtaining sizable collections of data set. Instead, for drawing classification and fake detection we relay only on one-vs-all settings.

\begin{table}[htp]
\caption{Validation of Stroke Classifier - Across Techniques ( 5 Classes: Picasso, Matisse, Schiele, Modigliani, Others) - five-folds mean (std)}
\begin{center}
\scalebox{1}{
\begin{tabular}{|l|l|l|}
\hline
Approach & Train & Test \\
\hline
Hand-Crafted - SVM &  55.01 \% (1.41\%) & 48.97\% (5.82\%)  \\
GRU						& 87.72\%	(2.43\%) & 	74.65\% (3.41\%) \\
\hline
\end{tabular}
}
\end{center}
\label{T:MultiClass}
\end{table}

\subsection{Drawing Classification and Detection of Fakes}
\label{S:ValidateDraw}

\noindent{\em Drawing Classification Validation:}  
Given the trained stroke classifiers, their performance is tested on drawing classification settings, also using one-vs-all settings. We used the four aforementioned strategies for aggregating the results from the stroke level to the drawing level. Given that the stroke classifiers are trained on a five-fold cross-validation setting, the drawing classification followed that strategy, i.e. in each fold, each drawing in the test split is classified using the classifier trained on the 80\% of the images in the training split, hence there is no standard deviation to report. Table~\ref{T:Drawing} shows the results for the across-technique setting and Table~\ref{T:DrawingTS} shows the results for the technique-specific setting.

\noindent{\em Evaluation on Fake Drawings:}
The trained stroke classifiers were also tested on the collected fake drawings to evaluate whether the classifiers are really capturing artists' stroke characteristics and invariants or just statistics that can be easily deceived by forged versions. We used the Picasso-vs-all stroke classifiers to test the fake drawings that are made to imitate Picasso drawings (We denote them as Picasso fakes). A similar setting is used for Matisse fakes and Schiele fakes. Since the stroke classifiers are trained on a five-fold setting, we have five different classifiers trained per artist, one for each fold. Each test stroke is classified using the five classifiers and the majority vote is computed. The different aggregation methods are used to achieve a final classification for each drawing.  Since one-vs-all setting is adapted, classifying a fake Picasso as others in a Picasso-vs-all setting is considered a correct classification, while classifying fake Picasso as Picasso is considered a wrong prediction. The bottom parts of Table~\ref{T:Drawing} and Table ~\ref{T:DrawingTS} shows the classification results for the fake dataset for the across-technique and technique-specific settings respectively. 

The table shows that the trained one-vs-all stroke classifiers for all the three artists, are robustly rejecting fake drawing with accuracy reaching 100\% in the across-technique case. 
A notable difference here is that the GRU failed to detect the fake drawings, in particular for the Picasso-vs-all, while the hand-crafted features detected all the fakes. Similar case happens for Schiele-vs-all as well. We hypothesize that this is because of the limited training data in the technique-specific case, which did not allow the GRU to learn an invariant model that generalizes well as in the across-technique case. In contrast the hand-crafted models did not suffer from this limitation. Overall, the hand-crafted features are outperforming in detecting the fakes.

\begin{table*}[tbh]
\caption{Validation of Drawing Classifiers -  - One-vs-All -{\bf Across Techniques}}
\begin{center}
\scalebox{0.7}{
\begin{tabular}{|l|l|l|l|l|l|l|l|l|l|}
\hline
\multicolumn{10}{|c|}{Across-Techniques } \\
\hline
 & \multicolumn{3}{|c|}{Picasso-vs-All} & \multicolumn{3}{|c|}{Matisse-vs-All} & \multicolumn{3}{|c|}{Schiele-vs-All} \\
 \hline
Aggregation & Hand-crafted & GRU & Combined  & Hand-crafted & GRU & Combined & Hand-crafted & GRU & Combined  \\
\hline
Majority  & 66.67\%	&		76.77\%	&	82.49\%	&	54.88\%	&	81.14\%	&	80.47\%	&	74.41\%	&	82.49\%	&	81.82\% \\
Posterior & 67.68\%&			77.44\%&		81.48\%&		56.90\%&		81.48\%&		79.12\%&		74.75\%&		83.50\%&		82.49\% \\
85\%-certain & 73.06\% &			79.80\%	&	82.83\%&		38.05\%&		80.47\%&		78.79\%&		75.42\%&		83.50\%&		83.84\% \\
Certainty-weighted & 67.34\%&			79.80\%&		82.83\%&		58.25\%&		80.81\%&		80.47\%&		75.42\%&		85.19\%&		83.16\% \\

\hline
\multicolumn{10}{|c|}{\bf Detection of Fake Drawings} \\
\hline
 & \multicolumn{3}{|c|}{Picasso-vs-All} & \multicolumn{3}{|c|}{Matisse-vs-All} & \multicolumn{3}{|c|}{Schiele-vs-All} \\
 \hline
Aggregation & Hand-crafted & GRU & Combined  & Hand-crafted & GRU & Combined & Hand-crafted & GRU & Combined  \\
\hline
Majority  & 100\%  &  87.50 \% & 100\%&  76.92 \% & 100\%  & 100\%&  100\%  & 100\%& 100\%\\
Posterior & 100\% & 87.50 \% & 100\%&  76.92 \% & 100\%  & 100\%& 100\%  & 100\%& 100\%\\
85\%-certain & 100\% & 87.50 \% & 100\%& 76.92 \% & 100\%  & 100\%& 100\%  & 100\%& 100\%\\
Certainty-weighted &  100\% & 87.50 \% & 100\%& 76.92 \% & 100\%  & 100\%& 100\%  & 100\%& 100\%\\
\hline
\end{tabular}
}
\end{center}
\label{T:Drawing}
\end{table*}

\begin{table*}[tbh]
\caption{Validation of Drawing Classifiers - One-vs-All - {\bf Technique-specific}}
\begin{center}
\scalebox{0.7}{
\begin{tabular}{|l|l|l|l|l|l|l|l|l|l|}
\hline
\multicolumn{10}{|c|}{Technique-Specific } \\
\hline
 & \multicolumn{3}{|c|}{Picasso-vs-All} & \multicolumn{3}{|c|}{Matisse-vs-All} & \multicolumn{3}{|c|}{Schiele-vs-All} \\
 \hline
Aggregation & Hand-crafted & GRU & Combined  & Hand-crafted & GRU & Combined & Hand-crafted & GRU & Combined  \\
\hline
Majority  & 	72.41\% &			82.76\% &		81.38\% &		65.52\% &		78.62\% &		82.76\% &		81.25\% &		78.12\% &		81.25\%  \\
Posterior &	72.41\% &			82.76\% &		81.38\% &		66.21\% &		79.31\% &		80.69\% &		84.38\% &		78.12\% &		81.25\% \\
85\%-certain &	72.41\% &			82.76\% &		82.76\% &		69.66\% &		76.55\% &		80.69\% &		84.38\% &		78.12\% &		81.25\%  \\
Certainty-weighted &	71.72\% &			82.76\% &		82.07\% &		69.66\% &		77.93\% &		80.00\% &		87.50\% &		78.12\% &		81.25\% \\


\hline
\multicolumn{10}{|c|}{\bf Detection of Fake Drawings} \\
\hline
 & \multicolumn{3}{|c|}{Picasso-vs-All} & \multicolumn{3}{|c|}{Matisse-vs-All} & \multicolumn{3}{|c|}{Schiele-vs-All} \\
 \hline
Aggregation & Hand-crafted & GRU & Combined  & Hand-crafted & GRU & Combined & Hand-crafted & GRU & Combined  \\
\hline
majority &	100.00\% &		12.50\% &	16.67\% &	94.87\% &	100.00\% &	100.00\% &	100.00\% &	45.00\% &	55.00\% \\
Posterior &	100.00\% &		12.50\% &	16.67\% &	97.44\% &	100.00\% &	100.00\% &	100.00\% &	45.00\% &	55.00\% \\
k-certain &	100.00\% &		12.50\% &	20.83\% &	97.44\% &	100.00\% &	100.00\% &	100.00\% &	45.00\% &	60.00\% \\
certainity-weighted	 & 100.00\% &		12.50\% &	20.83\% &	97.44\% &	100.00\% &	100.00\% &	100.00\% &	45.00\% &	60.00\% \\

\hline
\end{tabular}
}
\end{center}
\label{T:DrawingTS}
\end{table*}

\section{Conclusion}
In this paper we proposed an automated method for quantifying the characteristics of artist stokes in drawings. The approach is inspired by the Pictology methodology proposed by van Dantzig. The approach segments the drawing into individual strokes using a novel segmentation algorithm. The characteristics of each stroke is captured using global and local shape features as well as a deep neural network that captures the local shape and tone variations of each stroke. We compared different types of features and showed results at the stroke classification and drawing classification levels. 

The main result of this paper is that it shows that we can discriminate between artists at the stroke-level with high accuracy, even using images of drawing of typical off-the-web or scanned books resolutions. We also tested the methodology using a collected data set of fake drawings and the results show that the proposed method is robust to such imitated drawings, which highlights that the method can indeed capture artists' invariant characteristics  that is hard to imitate.

{\small
\bibliographystyle{unsrt}
\bibliography{Pictologyreferences}
}

\clearpage

\section*{Appendix: Ablation studies and other experiments}
\appendix

\subsubsection*{Adaptive GRU Patch Radius}

Table~\ref{T:Adaptive} shows a comparison between choosing the patch size based on an adaptive radius vs. fixed radius for the GRU model. For the fixed radius case, 11x11 patches are used. For the adaptive case,  a radius $r$ is computed for each drawing by computing the mean rip length for each stroke and taking the median over all strokes in the drawing. Square patches of size $2*r+1$ are used and scaled to 11x11 patches. The comparison in the table shows that the adaptive radius does not improve over the fixed radius in most of the cases. The comparison is shown for both the across-techniques and technique-specific cases. The adaptive radius showed improvement only in the case of technique-specific Schiele vs. all classification, 

\begin{table*}[tbh]
\caption{Comparison of GRU Drawing Classifiers using Adaptive vs. Fixed Radius Patches }
\begin{center}
\scalebox{0.8}{
\begin{tabular}{|l|l|l|l|l|l|l|}
\hline
\multicolumn{7}{|c|}{\bf Across-Techniques } \\
\hline
\multicolumn{7}{|c|}{one-vs-all drawing classification } \\
\hline
 & \multicolumn{2}{|c|}{Picasso-vs-All} & \multicolumn{2}{|c|}{Matisse-vs-All} & \multicolumn{2}{|c|}{Schiele-vs-All} \\
 \hline
Aggregation & Adaptive Radius & Fixed Radius  & Adaptive Radius & Fixed Radius  & Adaptive Radius & Fixed Radius  \\
\hline
Majority  &  78.11\% &			76.77\% &				78.79\% &		81.14\% &				73.40\% & 82.49\% \\
Posterior & 78.11\% &			77.44\% &				78.45\% &		81.48\% &				74.41\% & 83.50\% \\
85\%-certain & 79.46\% &			79.80\% &				76.77\% &		80.47\% &				76.77\% & 83.50\% \\
Certainty-weighted & 79.46\% &			79.80\% &				78.45\% &		80.81\% &				77.78\% &85.19\% \\
\hline
\hline
\multicolumn{7}{|c|}{\bf Technique-Specific } \\
\hline
\multicolumn{7}{|c|}{one-vs-all drawing classification } \\
\hline
 & \multicolumn{2}{|c|}{Picasso-vs-All} & \multicolumn{2}{|c|}{Matisse-vs-All} & \multicolumn{2}{|c|}{Schiele-vs-All} \\
 \hline
Aggregation & Adaptive Radius & Fixed Radius  & Adaptive Radius & Fixed Radius  & Adaptive Radius & Fixed Radius  \\
\hline
Majority  & 77.24\% &			82.76\% &				75.17\% &		78.62\% &				90.62\% &		78.12\% \\	
Posterior & 77.24\% &			82.76\% &				74.48\% &		79.31\% &				90.62\% &		78.12\% \\	
85\%-certain & 77.24\% &			82.76\% &				76.55\% &		76.55\% &				90.62\% &		78.12\% \\	
Certainty-weighted & 78.62\% &			82.76\% &				77.24\% &		77.93\% &				87.50\% &		78.12\% \\
\hline
\hline
\end{tabular}
}
\end{center}
\label{T:Adaptive}
\end{table*}

\subsubsection*{Ablation study of the Hand-crafted Features}
In this set of experiments we conduct an ablation study of the elements of the hand-crafted stroke features. These experiment is done on a binary classification setting to discriminate between the strokes of Picasso and Matisse drawn using ink/pen technique. SVM with polynomial kernel of degree 3 is used in all experiments. Five-fold cross validation is performed.

\begin{table*}[tbh]
\caption{Ablation study of the Hand-crafted features - five-folds}
\begin{center}
\scalebox{0.9}{
\begin{tabular}{|l|l|l|l|l|}
\hline
Feature & \multicolumn{2}{|c|}{training accuracy} & \multicolumn{2}{|c|}{test accuracy} \\
&  mean & std & mean & std \\
\hline
Fourier Descriptors (FD)  & 	61.75\% &	1.11\% &	57.55\% &	5.46\% \\
Reconstruction Error Profile (REP)  &	64.39\% &	0.75\% &	63.20\% &	11.09\% \\
Stroke Thickness Profile (STP)	& 74.11\% &	1.33\% &	63.68\% &	2.58\% \\
Curvature	 & 66.22\% &	0.60\% &	60.17\% &	2.06\% \\
Stroke Length (SL) &	57.30\% &	0.28\% &	57.12\% &	1.86\% \\
FD+REP	& 69.23\% &	0.57\% &	64.23\% &	6.69\% \\
STP + Curvature	& 79.63\% &	0.71\% &	68.93\% &	4.04\% \\
STP + Curvature + SL	& 80.60\% &	0.80\% &	70.17\% &	3.81\% \\
FD+REP+STP	& 77.42\% &	1.95\% &	68.35\% &	4.58\% \\
FD+REP+STP+Curvature &	85.91\% &	0.88\% &	74.13\% &	6.47\% \\
FD+REP+STP+Curvature+SL &	86.70\% &	1.08\% &	75.07\% &	5.67\%  \\
\hline
\end{tabular}
}
\end{center}
\label{T:Adaptive}
\end{table*}

\clearpage

Figure~\ref{F:Fakes} key: \\
Fake, Fake, Matisse \\
Matisse, Fake, Fake, Matisse \\
Fake, Matisse, Picasso, Fake \\
Fake, Picasso, Picasso, Fake \\
Schiele, Fake, Fake, Schiele, Schiele, Fake \\

\section*{About the Authors:}

\begin{tabular}{ll}
\includegraphics[width=1in]{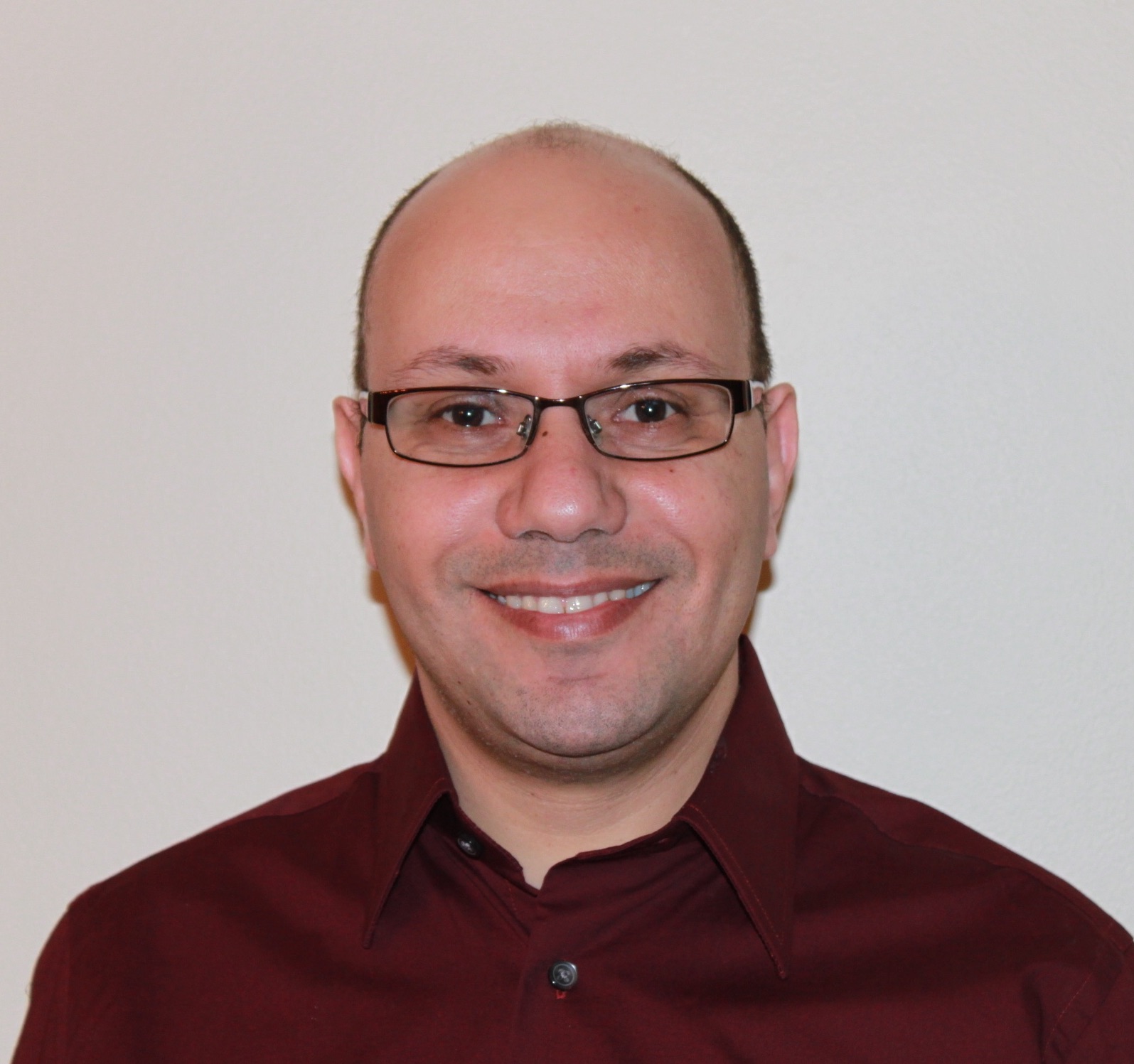} &
\parbox{5in}{{\bf Dr. Ahmed Elgammal:} is a professor at the Department of Computer Science, Rutgers University. He is the founder and director of the Art and Artificial Intelligence at Rutgers, which focuses on data science in the domain of digital humanities. He is also an Executive Council Faculty at Rutgers University Center for Cognitive Science. Prof. Elgammal has published over 160 peer-reviewed papers, book chapters, and books in the fields of computer vision, machine learning and digital humanities. He is a senior member of the Institute of Electrical and Electronics Engineers (IEEE). He received the National Science Foundation CAREER Award in 2006.  Dr. Elgammal recent research on knowledge discovery in art history received wide international media attention, including reports on the Washington Post, New York Times, NBC News, the Daily Telegraph, Science News, and many others. Dr. Elgammal is the founder and CEO of Artrendex, a startup that builds innovative AI technology for the art market.  Dr. Elgammal received his M.Sc. and Ph.D. degrees in computer science from the University of Maryland, College Park, in 2000 and 2002, respectively.
} \\
\includegraphics[width=1in]{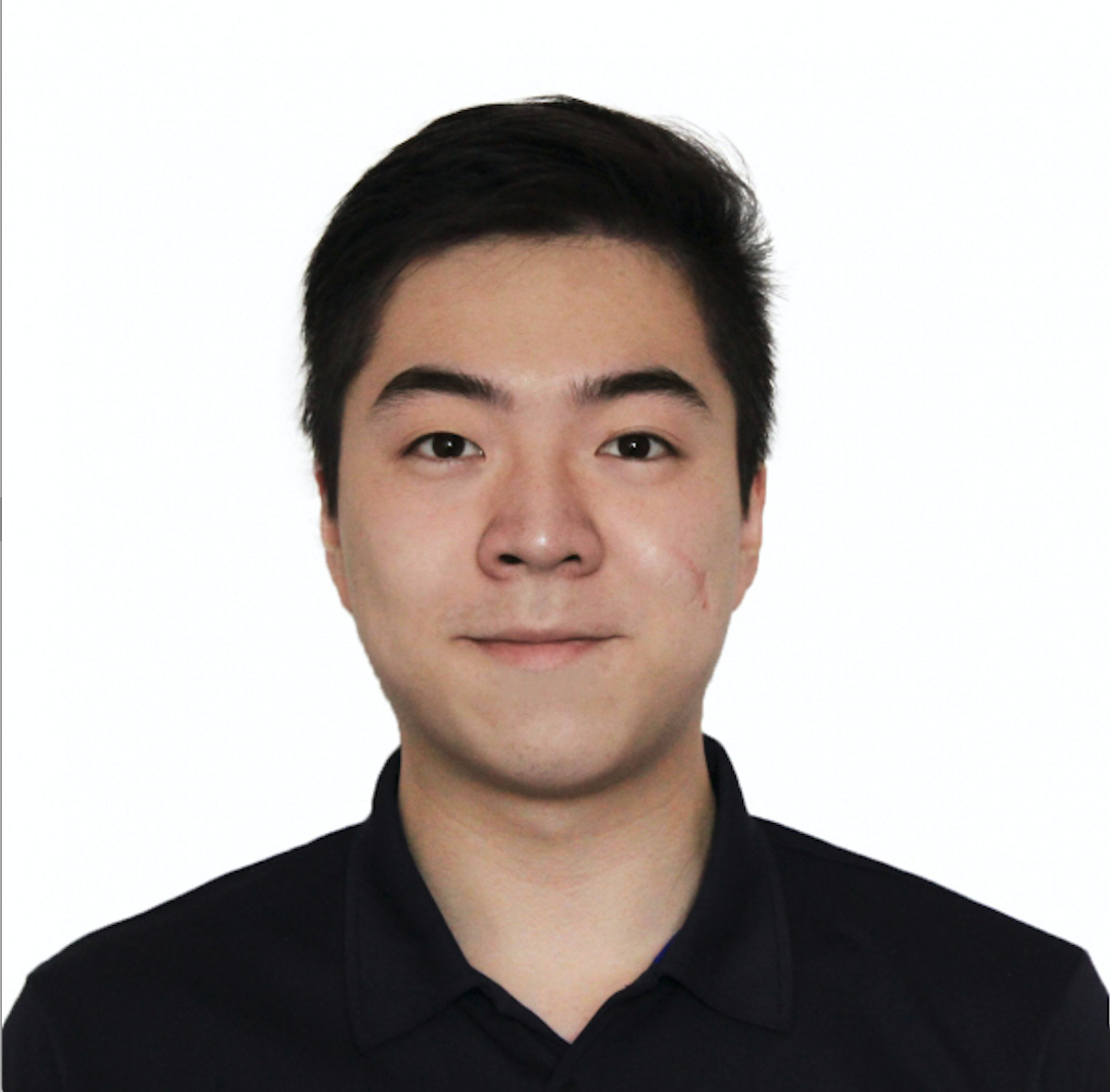} & 
\parbox{5in}{ {\bf Yan Kang} main research interests are Deep Learning,Computer Vision and Natural Language Processing. Mr. Kang was a senior deep learning researcher at Artrendex. 
Currently Mr. Kang is  a Computer Vision Engineer at JD.com American Technologies Corporation, working on Object Classification \& Detection and Neural Network Compression. He obtained his M.S. degree in Computer Science at Rutgers University (New Brunswick) in May 2016.  
} \\
\end{tabular}

\begin{tabular}{ll}
\includegraphics[width=1in]{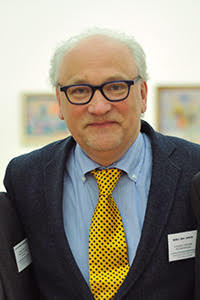} &
\parbox{5in}{
{\bf Milko den Leeuw} is a painting conservator, specialized in the technical and scientific investigation of paintings. He completed his training in painting conservation and pictology (an analytical method for attribution and evaluation of paintings) at the studio Dora van Dantzig in Amsterdam in 1989. In 1991 he founded the Atelier for Restoration \& Research of Paintings (ARRS) with drs Ingeborg de Jongh.
ARRS was from 1991-2006 the outdoor conservator atelier of the Rijksmuseum Catharijneconvent Utrecht. ARRS has produced numerous international publications concerning new techniques in authentication research and rediscoveries of paintings. ARRS has been involved in the conservation of important masters in Dutch Fine Art such as Rembrandt van Rijn, Hendrick ter Brugghen and Gerard ter Borch and has contributed to the rediscovery of paintings by Jan Lievens, Capser Netscher and Quinten Matsijs. ARRS has also performed conservation and research on National Art Treasures like the ceiling dating from 1672 by Gerard de Lairesse from the International Court for Peace \& Justice in The Hague (Carnegie Foundation), and Italian masters Giovanni Bellini and Canaletto, French masters Manet and Degas, German masters Lucas Cranach and Max Beckmann, Russian masters as Kasimir Malevich and Nathalia Goncharova and American masters Andy Warhol and Mark Rothko. ARRS is a professional equipped lab for conservation treatment and (authentication) research on paintings. ARRS holds a worldwide reputation for bridging art history, conservation technique and material sciences.
ARRS is Authentication in Art organizer of the 2012, 2014, 2016 AiA conferences.} \\
\end{tabular}

\end{document}